\documentclass[useAMS]{mn2e}
\usepackage{amsmath}
\usepackage{textcomp}
\usepackage{amssymb}
\usepackage[pdftex]{graphicx}
\usepackage{amssymb}

\usepackage[margin=.8in, paperwidth=8.5in, paperheight=11in]{geometry}
\usepackage{epsfig}
\usepackage{graphicx}
\usepackage{multirow}
\usepackage{multicol}
\usepackage{caption}
\usepackage{subfig}
\usepackage{esint}

\begin{document}

\title{Non-Axisymmetric Line Driven Disc Winds II - Full Velocity Gradient}
\author[S. Dyda, D. Proga]
{\parbox{\textwidth}{Sergei~Dyda\thanks{sdyda@physics.unlv.edu}, Daniel Proga}\\
Department of Physics \& Astronomy, University of Nevada Las Vegas, Las Vegas, NV 89154 
}

\date{\today}
\pagerange{\pageref{firstpage}--\pageref{lastpage}}
\pubyear{2017}

\label{firstpage}

\maketitle

\begin{abstract}
We study non-axisymetric features of 3D line driven winds in the Sobolev approximation, where the optical depth is calculated using the full velocity gradient. We find that non-axisymmetric density features, so called clumps, form primarily at the base of the wind on super-Sobolev length scales. The density of clumps differs by a factor of $\sim 3$ from the azimuthal average, the magnitude of their velocity dispersion is comparable to the flow velocity and they produce $\sim 20\%$ variations in the column density. Clumps may be observable because differences in density produce enhancements in emission and absorption profiles or through their velocity dispersion which enhances line broadening.                
 
\end{abstract}

\section{Introduction}
Many accretion disc systems such as Active Galactic Nuclei (AGN) and cataclysmic variables (CV) are observed to have blue shifted absorption lines, which are interpreted as evidence for disc winds. A possible driving mechanism for launching these winds is radiation pressure from spectral lines, so called line driving. Line driving has been a well studied driving mechanism for outflows since seminal work by Lucy and Solomon (1970) and Castor, Abbott and Klein (1975, hereafter CAK) applied it to stellar winds from OB stars. We describe many of the relevant theoretical and numerical studies of line driven winds in the introduction of our companion paper Dyda \& Proga 2018 (hereafter DP18).

Line driven winds can exhibit fine scale structure or clumps on multiple length scales. On sub-Sobolev length scales, clumping can occur due to the line deshadowing instability (LDI) where small amplitude waves in the subsonic part of the flow are amplified by the line driving into non-linear waves in the supersonic part of the flow. This has been demonstrated in both 1D (Owocki, Castor \& Rybicki 1988, hereafter OCR) and recently 2D (Sundqvist, Owocki \& Puls 2017, hereafter SOP17) models, where typical clump sizes are found to be of order the Sobolev length. This is a leading order effect for generating clumps in spherical flows, as they are otherwise smooth on larger length scales.

For disc winds, on super-Sobolev scales, clumps can be generated because geometric effects cause flows to be unsteady. Line accelertion for 1D radial flows accelerated by a radiating point source, as introduced by CAK, is a function of only the radial velocity gradient. For geometrically more complex flows and radiation fields, this treatment is modified (see Owocki, Cranmer \& Gayley 1996, hereafter OCG96) and requires computing the velocity gradient along lines of sight from the radiating surface to the accelerating gas. In the earliest axisymmetric simulations of line driven disc winds, Proga, Stone \& Drew (1998, hereafter PSD98) approximated the full velocity gradient as the radial and vertical gradients for stellar and disc radiation, respectively. They found outflows were non-stationary and had density structures at the base of the wind. Their approach notably neglected azimuthal terms in the velocity gradient that are present for rotating flows. Subsequent work by Proga, Stone and Drew (1999, hereafter PSD99) included all terms in the velocity gradient, the so called \emph{full-Q} treatment. They found including these additional terms did not significantly alter the wind solutions and demonstrated that for disc winds primarily driven by radiation from the disc that non-stationarity is a robust feature of this type of outflow.

Relaxing axisymmetry allows clumps to form in the axial direction and enhances the density structures that form at the base of the wind. In DP18, we approximated the velocity gradient as in PSD98 and found that in 3D an initial vertical subsonic disc perturbation induces density structures to form at the base of the wind that differ from the azimuthal average by a factor of a few. We explicitly used a non-axisymmetric initial condition and showed that line driving does not cause these non-axisymmetric structures to grow without bounds. In the present work, as in PSD99, we include all terms in the velocity gradient tensor. The inclusion of azimuthal gradient terms, owing to disc rotation, means that the line driving force has non-axisymmetric components and axial, super-Sobolev length, clumps can grow from axisymmetric initial conditions.    

Clumpiness can affect the flow by altering the ionization structure of the wind, via shielding outer parts of the flow (see for example Fullerton 2011). This is a possible mechanism for coupling the inner, subsonic, flow with the outer, supersonic wind. Emission features such as IR continuum and subordinate lines like $H\alpha$ are the result of two body processes and therefore sensitive to density squared. Line absorption, dependent on density, is also sensitive, albeit less so, to the clumps. Clumping in disc winds has been used for instance to explain in part line width ratios seen in BAL QSOs (Matthews et al. 2016). Velocity dispersion of clumps may enhance line broadening and observationaly might mimic or be misinterpreted as turbulence.

The purpose of this work is to explore the properties of non-axisymmetries in 3D line driven disc winds from azimuthal terms in the velocity gradient tensor. We apply this to CV systems where we assume the radiation field is provided by a spherical central object and a re-radiating thin disc (Shakura \& Sunyaev 1973). We work in the Sobolev approximation and explore cases for different relative luminosity between the disc and central star. As in axisymmetric simulations, we show that inclusion of azimuthal terms in the velocity gradient tensor do not affect the overall character of the disc wind. However, we show that significant axial structures form, which are otherwise absent. Non-axisymmetric effects are stronger than when induced via a $\phi-$dependent initial conditions as in DP18.   

In Section \ref{sec:numerical}, we describe our numerical methods, in particular our treatment of the radiation force. In Section \ref{sec:results}, we describe the main features of our wind solutions and characterize the resulting non-axisymmetries in terms of density correlation lengths, velocity dispersion and Fourier modes. In Section \ref{sec:discussion}, we comment on possible observational diagnostics. We summarize our work and discuss possible future directions of inquiry in Section \ref{sec:conclusion}.

\section{Numerical Methods}
\label{sec:numerical}
We performed all numerical simulations with the publicly available MHD code \textsc{Athena++} (Gardiner \& Stone 2005, 2008). The basic physical setup is a gravitating, luminous, central object surrounded by a thin, luminous accretion disc. Both the central object and accretion disc source a radiation field to drive the gas that is optically thin to the continuum.  Our model is decribed in detail in a companion paper DP18, but we sketch the basic setup here, as well as any differences from our previous work below. 

\subsection{Basic Equations}
\label{sec:basic_equations}
The basic equations for single fluid hydrodynamics driven by a radiation field are
\begin{subequations}
\begin{equation}
\frac{\partial \rho}{\partial t} + \nabla \cdot \left( \rho \mathbf{v} \right) = 0,
\end{equation}
\begin{equation}
\frac{\partial (\rho \mathbf{v})}{\partial t} + \nabla \cdot \left(\rho \mathbf{vv} + \mathbf{P} \right) = - \rho \nabla \Phi + \rho \mathbf{F}^{\rm{rad}},
\end{equation}
\begin{equation}
\frac{\partial E}{\partial t} + \nabla \cdot \left( (E + P)\mathbf{v} \right) = -\rho \mathbf{v} \cdot \nabla \Phi + \rho \mathbf{v} \cdot \mathbf{F}^{\rm{rad}} ,
\label{eq:energy}
\end{equation}
\label{eq:hydro}%
\end{subequations}
where $\rho$, $\mathbf{v}$ are the fluid density and velocity respectively and $\mathbf{P}$ is a diagonal tensor with components P the gas pressure. For the gravitational potential, we use $\Phi = -GM/r$ and $E = 1/2 \rho |\mathbf{v}|^2 + \mathcal{E}$ is the total energy where $\mathcal{E} =  P/(\gamma -1)$ is the internal energy. The total radiation force per unit mass is $\mathbf{F}^{\rm{rad}}$. The isothermal sound speed is $a^2 = P/\rho$ and the adiabatic sound speed $c_s^2 = \gamma a^2$. We take a nearly isothermal equation of state $P = k \rho^{\gamma}$ where $\gamma = 1.01$.   We can compute the temperature from the internal energy via $T = (\gamma -1)\mathcal{E}\mu m_{\rm{p}}/\rho k_{\rm{b}}$ where $\mu = 0.6$ is the mean molecular weight and other symbols have their usual meaning.

\subsection{Radiation Force}
\label{sec:radiation_force}
We assume a time independent radiation field is sourced by a spherically symmetric star at the origin and an axisymmetric, geometrically thin accretion disc along the midplane. The frequency integrated intensity of the star is
\begin{equation}
I_* = \Gamma_* \frac{GM}{\pi r_*^2} \frac{c}{\sigma_e},
\end{equation}
where $r_*$ is the radius of the star, $\sigma_e$ is the Thompson cross section per unit mass, $c$ the speed of light, and $\Gamma_*$ the stellar Eddington fraction. The frequency integrated intensity of the disc is
\begin{align}
I(r_d) &= \frac{3}{\pi} \frac{GM}{r_*^2}\frac{c}{\sigma_e} \Gamma_d \Bigg\{ \left( \frac{r_*}{r_d} \right)^3 \left[ 1 - \left(\frac{r_*}{r_d} \right)^{1/2}\right]  \\ &+  \frac{x}{3\pi} \left[ \sin^{-1} \left( \frac{r_*}{r_d}\right)  - \frac{r_*}{r_d} \left( 1 - \left(\frac{r_*}{r_d} \right)^{2} \right)^{1/2} \right] \Bigg\}, 
\end{align}  
where $x = \Gamma_*/\Gamma_d$ is the ratio of stellar to disc luminosity and the disc Eddington number
\begin{equation}
\Gamma_d = \frac{\dot{M}_{\rm{acc}} \sigma_e}{8 \pi c r_*},
\end{equation}
where $\dot{M}_{\rm{acc}}$ is the accretion rate in the disc (e.g., Pringle 1981). The disc and central star are taken to be optically thick. 

We assume all gas is optically thin to continuum radiation and every point in the wind experiences a radiation force
\begin{equation}
\mathbf{F}^{\rm{rad}} = \mathbf{F}^{\rm{rad}}_e + \mathbf{F}^{\rm{rad}}_{L},
\end{equation}  
which is a sum of the contributions due to electron scattering $\mathbf{F}^{\rm{rad}}_e$ and line driving $\mathbf{F}^{\rm{rad}}_{L}$. In this continuum optically thin approximation, the radiation force due to electron scattering is
\begin{equation}
\mathbf{F}^{\rm{rad}}_e = \varoiint \left( \mathbf{n} \frac{\sigma_e I d\Omega}{c} \right),
\end{equation} 
where $\mathbf{n}$ is the normal vector from the radiating surface to the point in the wind, $d\Omega$ is the solid angle and the integration is carried out over the entire disc and star. 

We treat the radiation due to lines using a modification of the CAK formulation where the radiation force due to lines is 
\begin{equation}
\mathbf{F}^{\rm{rad}}_L = \varoiint M(t) \left( \mathbf{n} \frac{\sigma_e I d\Omega}{c} \right),
\label{eq:f_line}
\end{equation}
and $M(t)$ is the so-called force multiplier. The force multiplier parametrizes how many lines are effectively available to increase the scattering coefficient. We use the OCR parametrization of the line strength, where working in the Sobolev approximation, it is a function of the optical depth parameter
\begin{equation}
t = \frac{\sigma_e \rho v_{\rm{th}}}{|dv_{l}/dl|},
\label{eq:optical_depth_parameter}
\end{equation} 
where $v_{\rm{th}}$ is the thermal velocity of the gas and $dv_{l}/dl$ is the velocity gradient along the line of sight. The OCR formulation for the force multiplier is
\begin{equation}
M(t) = k t^{-\alpha} \left[ \frac{(1 + \tau_{\rm{max}})^{1-\alpha} - 1}{\tau_{\rm{max}}^{1 - \alpha}}\right],
\label{eq:force_multiplier}
\end{equation} 
where $k$ and $\alpha$ are constants, $\tau_{\rm{max}} = t \eta_{\rm{max}}$, and $\eta_{\rm{max}}$ is related to the maximum force multiplier via $M_{\rm{max}} = k (1 - \alpha) \eta_{\rm{max}}^{\alpha}$. This force multiplier parametrization follows CAK for large optical depths but saturates to $M_{\rm{max}}$ as optical depth becomes very small.

The velocity gradient can be expressed as a sum over elements of the shear tensor $\epsilon_{ij}$ via 
\begin{equation}
\frac{dv_l}{dl} \equiv Q = \epsilon_{ij} n_i n_j,
\label{eq:fullQ}
\end{equation}
where $n_i$ are the components of the normal vector. In this work, unlike in DP18, we use all terms appearing in equation (\ref{eq:fullQ}), which we refer to as the \emph{full-Q} formalism. Additional details about our numerical treatment of the radiation force can be found in the Appendix of DP18.

\subsection{Simulation Parameters}
\label{sec:simulation_parameters}
We chose parameters that correspond to the simulations of PSD99 and generally followed their setup as closely as possible given that they used the \textsc{Zeus} 2D code (Stone \& Norman 1992). The central object has mass and radius $M = 0.6 \ M_{\odot}$ and $r_* = 8.7 \times 10^{8} \ \rm{cm}$, respectively. The sound speed $c_s = 14 \ \rm{km/s}$, which corresponds to a hydrodynamic escape parameter $\rm{HEP} = GM/r_*c_s^2 = 4.6 \times 10^4$ at the base of the wind. For this high HEP, thermal driving, which requires $\rm{HEP}$ no more than $10$ (Stone \& Proga 2009; Dyda et al. 2017), is negligible throughout the domain.

The radial computational domain extends over the range $r_* \leq r \leq 10 \ r_*$ with logarithmic spacing $dr_{i+1}/dr_i = 1.05$. The polar angle range is $0 \leq \theta \leq \pi/2$ and has logarithmic spacing $d\theta_{j}/d\theta_{j+1} = 1.05$ which ensures that we have sufficient resolution near the disc midplane to resolve the density length scale, $\lambda_{\rho}$, of the hydrostatic equilibrium disc that forms i.e., to obtain $r \Delta \theta \ll \lambda_{\rho}$. We use linear spacing in the $\phi$ direction with range $0 \leq \phi \leq \pi/4$. We use a grid resolution $n_r \times n_{\theta} \times n_{\phi} = 96 \times 96 \times 64$ and $N_r \times N_{\theta} \times N_{\phi} = 3 \times 3 \times 2 = 18$ MPI meshblocks of size $32^3$. Meshblocks of size 32 is the recommended smallest size for \textsc{Athena++} to achieve good scaling with number of processors. In the poloidal plane, this resolution allows us to resolve the subsonic part of the flow. In the axial direction, we can resolve the $n_{\phi}/10 \sim 6$ highest frequency modes that can be excited in this wedge ( m = 8, 16, ...) . 

We impose outflow boundary conditions at the inner and outer radial boundaries and axis boundary conditions along the $\theta = 0$ axis. We assume a reflection symmetry about the $\theta = \pi/2$ midplane. In the $\phi$ direction, we impose periodic boundary conditions, where our 3D simulation covers a range $0 \leq \phi \leq \pi/4$.

Initially, the cells along the disc are set to have $\rho = \rho_d$, $v_r = v_{\theta} = 0$, $v_{\phi} = v_K$ and the pressure $P = c_s^2 \rho^{\gamma}$. In the rest of the domain $\rho = 10^{-20} \rm{g} \ \rm{cm}^{-3}$, $P = c_s^2 \rho^{\gamma}$ and $v_r = v_{\theta} = v_{\phi} = 0$.

After every full time step we reset $\rho_d$ to $10^{-8} \rm{g \ cm}^{-3}$, $v_r$ to $0$ and $v_{\phi}$ to $v_K = \sqrt{GM/r}$ along the disc. We also impose that the velocity $v_{\theta}$ is unchanged due to resetting density by resetting the momentum i.e. $(\rho v)_{\theta}^{n+1} = \rho_d (\rho v)_{\theta}^{n}/(\rho)^{n}$ where the superscripts refers to the $n$ and $n+1$ timestep respectively. Along the star, after every full timestep, we ensure the density $\rho_* \geq 10^{-16} \rm{g \ cm}^{-3}$. This density is low enough to be insignificant in contributing a stellar wind to the net outflow, but is needed to keep the density and velocity well behaved along the axis. 

We take the disc Eddington number $3.76 \times 10^{-4} \leq \Gamma_d \leq 1.18 \times 10^{-3}$. The thermal velocity of the gas is $v_{\rm{th}} = 4.2 \times 10^5 \rm{cm/s}$. The line driving parameters are chosen to be $k = 0.2$, $\alpha = 0.6$ and $M_{\rm{max}} = 4400$. The disc as a source of radiation is taken to extend outside the computational domain to $R_d = 30 \ r_*$.   

We impose a density floor $\rho_{\rm{floor}} = 10^{-22} \rm{g \ cm}^{-3}$, which ensures that $\rho \geq \rho_{\rm{floor}}$ throughout the domain, while conserving momentum. In addition, we implement a pressure floor $P_{\rm{floor}} = c_s^{2} \rho_{\rm{floor}}^{\gamma}$, so that our nearly isothermal condition is preserved.

\section{Results}
\label{sec:results}

\begin{table*}
\begin{center}
    \begin{tabular}{| c | c | c | c | c | c | c| c |c| c |}
    \hline \hline 
Model	& Q &	$\Gamma_d$	&x 	&\multicolumn{2}{c}{$\dot{M} \ [\rm{M_{\odot}/y}] $}	&\multicolumn{2}{c}{$v_r \ [\rm{km/s}]$}	&\multicolumn{2}{c}{$\omega \ [^{\circ}]$} \\
	&   & 	&  	& \textsc{Zeus-2D}	& \textsc{Athena++} & \textsc{Zeus-2D}	& \textsc{Athena++} & \textsc{Zeus-2D}	& \textsc{Athena++} \\ \hline
DP18u	& dv/dz & $3.76 \times 10^{-4}$	&  0	& $4.8 \times 10^{-14}$	& $1.77 \times 10^{-14}$ & 900	& 900 & 42	&  45\\
DP18p	& dv/dz &$3.76 \times 10^{-4}$	&  0	& 	& $2.14 \times 10^{-14}$ & 	& 900 & 	&  43\\ \hline
A	& Full  &$3.76 \times 10^{-4}$	&  0	& $5.5 \times 10^{-14}$	& $2.3 \times 10^{-13}$ & 900	& 1100 & 50	&  50\\
B	& Full  &$1.18 \times 10^{-3}$	&  0	& $4.0 \times 10^{-12}$	& $7.7 \times 10^{-12}$ & 3500	& 3300 & 60	&  62\\
C	& Full  &$1.18 \times 10^{-3}$	&  1	& $2.1 \times 10^{-11}$	& $2.4 \times 10^{-11}$ & 3500	& 3000 & 32	&  36\\
D	& Full  &$1.18 \times 10^{-3}$	&  3	& $7.1 \times 10^{-11}$	& $7.4 \times 10^{-11}$ & 5000	& 3500 & 16	&  18\\ \hline \hline

    \end{tabular}
\end{center}
\caption{Summary of global wind properties, mass flux $\dot{M}$, fast-stream velocity $v_r$ and wind opening angle $\omega$ for 3D \textsc{Athena++} models and comparison to analogous 2D axisymmetric simulations performed with \textsc{Zeus-2D}. DP18u and DP18p are the unperturbed and perturbed disc simulations respectivly from DP18, analogous to ``Run 2" in PSD98. New simulations in this work, Models ``A"..``D" share simulation parameters with models in PSD99.}
\label{tab:summary}
\end{table*}

\begin{figure}
                \centering
                \includegraphics[width=0.5\textwidth]{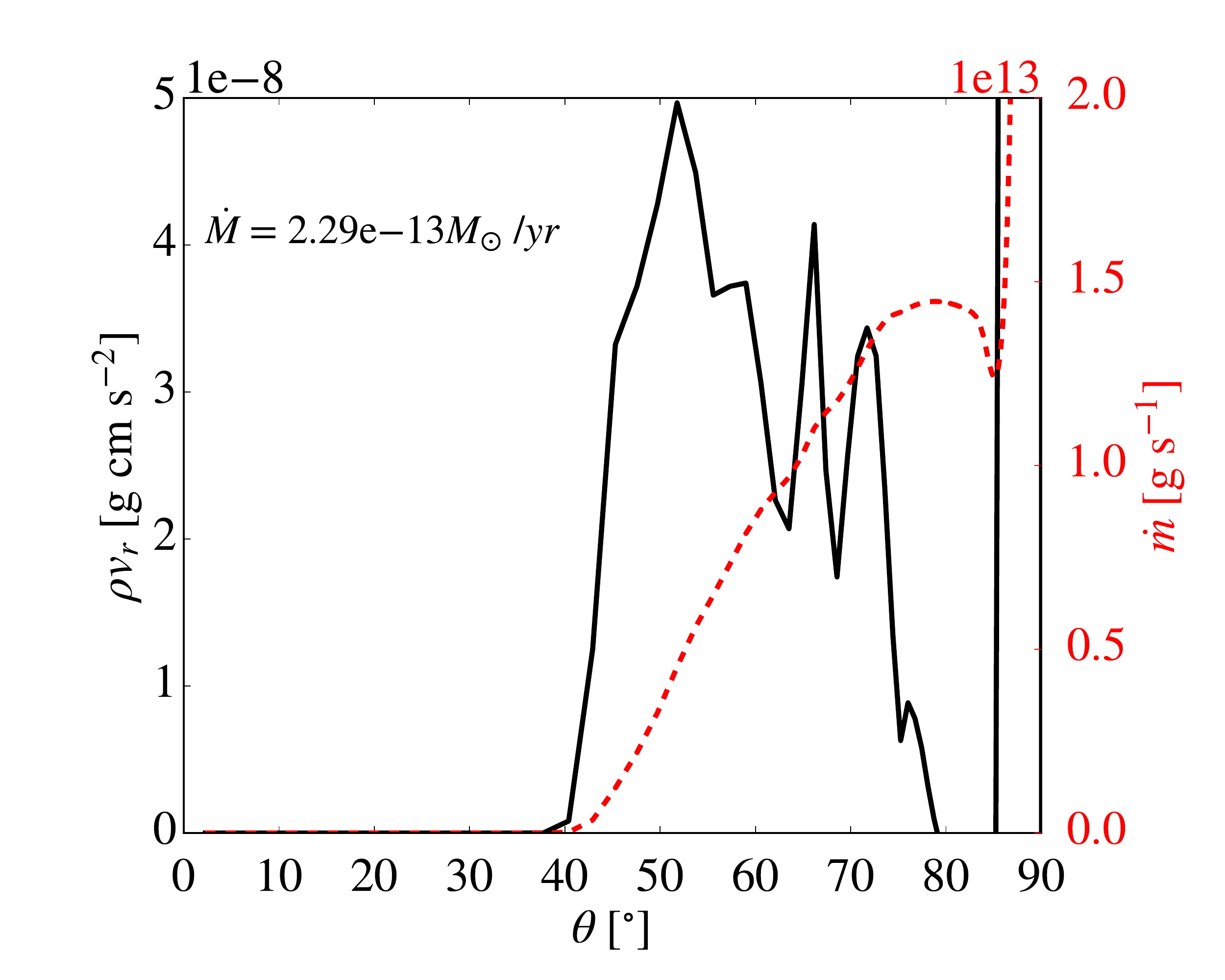}
                \includegraphics[width=0.5\textwidth]{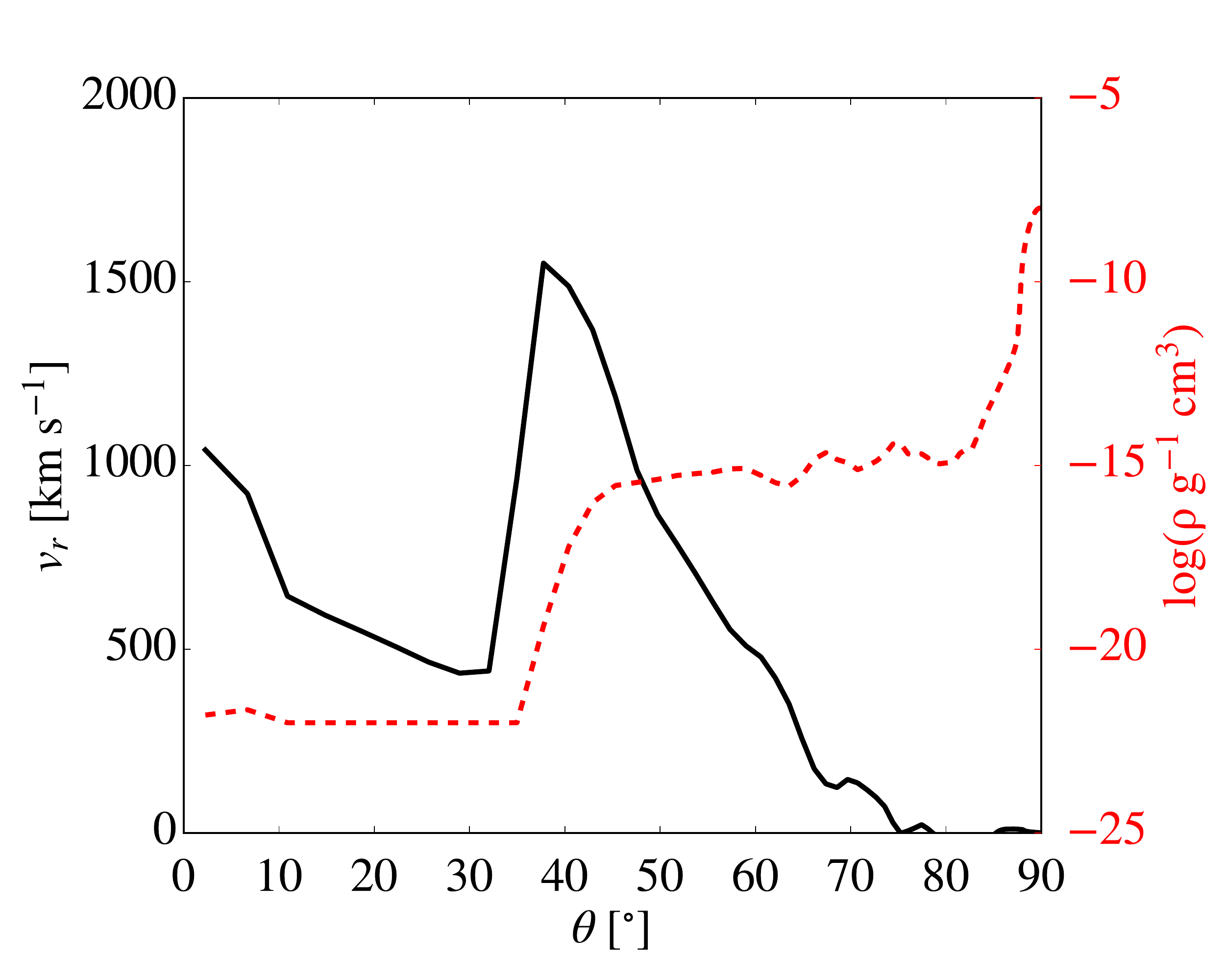}
        \caption{\textit{Top} - Momentum flux $\rho v_r$ (black, solid line) and integrated mass flux $\dot{m}$ (red, dashed line) at the outer boundary at $t = 1000$ s for Model A. \textit{Bottom} - Radial velocity $v_r$ (black, solid line) and density $\rho$ (red, dashed line) along the outer boundary.}
\label{fig:mdot_out_3d}
\end{figure} 

\begin{figure*}
                \centering
                \includegraphics[width=\textwidth]{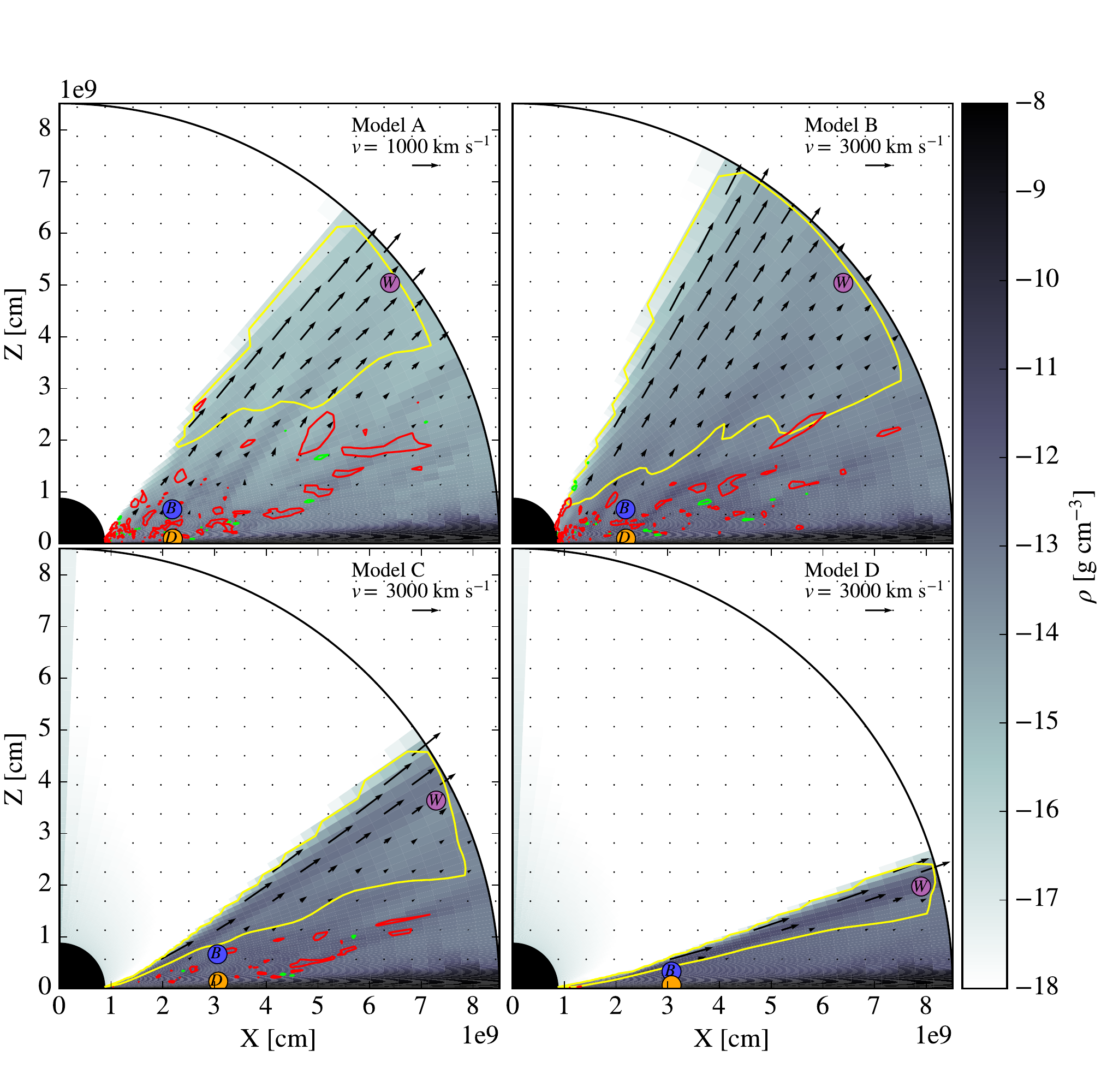}
        \caption{Plot of $\phi-$averaged density $\bar{\rho}$ (grayscale), overdensities (green), underdensities (red) and fast-stream contour (yellow) where $|\bar{\mathbf{v}}_p| = 1/3 v_{\rm{esc}}$ for Model A (top left), B (top right), C (bottom left) and D (bottom right) at t = 1000 s. We also indicate representative locations of the disc (orange), base (blue) and wind (purple) points.}
\label{fig:3D-Q_summary}
\end{figure*} 

We performed simulations for various disc and stellar luminosities. In Table \ref{tab:summary}, we summarize the global outflow properties of these solutions and compare them to 2D axisymmetric runs with the same parameters performed using \textsc{Zeus} in PSD99. We also include results from the unperturbed (u) and perturbed (p) solutions in DP18 and the analogous 2D axisymmetric \textsc{Zeus} run from PSD98.  

All winds are qualitatively similar. A fast-stream, containing most of the mass flux, is launched at an intermediate angle from the disc. Above this fast stream is a polar funnel region containing very low density gas. Between the fast-stream and the hydrostatic disc the flow is slower. We illustrate the properties of Model A in Fig \ref{fig:mdot_out_3d}, where we plot quantities along the outer boundary $r = r_{\rm{o}}$ at $t = 1000$ s. In the top panel, we plot the mass flux $\rho v_r$ (black, solid line) and the $\theta-$integrated mass flux (red, dashed line)
\begin{equation}
\dot{m}(\theta) = \int_0^{2\pi} \int_0^{\theta}    \rho(r_{\rm{o}},\theta',\phi) v_r(r_{\rm{o}},\theta',\phi)  \ \sin \theta' \, d\theta' d\phi,
\end{equation}  
In the lower panel, we plot the radial velocity $v_r$ (black, solid line) and the density $\rho$ (red, dashed line) at the outer boundary. The polar funnel is at $\theta \lesssim 40^{\circ}$ and the disc $\theta \gtrsim 85^{\circ}$. In Table \ref{tab:summary}, we quote the mass loss rate as $\dot{M} = \dot{m}(80^{\circ})$ to avoid fluctuations from the disc where density is high. Likewise the wind opening angle is quoted as $\omega = 50^{\circ}$ since this is the size of the wedge excluded from the polar funnel but including the disc.  

In Fig \ref{fig:3D-Q_summary}, we plot the $\phi-$averaged density $\bar{\rho}$ (grayscale) and velocity $\bar{\mathbf{v}}_p$ (black vectors) for each of our runs at t = 1000 s. We indicate locations of azimuthal over-densities (green lines), where $\rho_k \geq 3 \bar{\rho}$ and under-densities (red lines), where $\rho_k \leq \bar{\rho}/3$ and k is the $\phi$ index. We identify three qualitatively different regions in our simulations - a nearly hydrostatic \emph{disc}, a wind \emph{base} and the main part of the \emph{wind} (orange, blue, and purple points, respectively). We also indicate where the poloidal velocity is 1/3 the escape velocity, $|\bar{v}_{p}| = v_{\rm{esc}}/3$ with a yellow contour and take this to be the boundary of the fast-stream. We note that this region is devoid of over-dense and under-dense features.       

The mass flux and velocity of the fast-stream is controlled by the total system luminosity. Increasing the total luminosity of the star and disc, $\Gamma = \Gamma_* + \Gamma_D$, increases both the mass flux and the velocity of the fast-stream. The geometry of the flow is controlled by the relative luminosity of the star and the disc. Increasing the disc luminosity causes the flow to be more vertical whereas increasing the stellar luminosity causes the flow to be more radial. Winds driven by disc radiation ($x = 0$) are less stationary than winds driven by stellar radiation ( $x > 0$), the former having temporal variations of $\sim 2$ in their mass flux. 

These results are largely in agreement with results from PSD99. A noticable discrepancy is in the mass flux for Model A. To launch a wind the radiation force must overcome gravity. In the case of a saturated force multiplier, this requires a minimum luminosity $\Gamma_{\rm{min}}$ satisfying $(1 + x) \Gamma_{\rm{min}} M_{\rm{max}} = 1$. For Model A where x = 0, $\Gamma_{D} \gtrsim \Gamma_{\rm{min}} = 2.3 \times 10^{-4}$. Cases close to this minimum luminosity threshold are most sensitive to how each term of the force multiplier is calculated, particularly the optical depth, $t$, and thus the velocity gradient Q. The approximation $Q \approx |dv/dz|$ used in PSD98 and DP18 \emph{overestimates} $\tau_{\rm{max}}$ in equation (\ref{eq:force_multiplier}) and thus \emph{underestimates} the force multiplier. We verified this by running a case with $\Gamma_D = 3 \times 10^{-4}$ and found $\dot{M} = 4.7 \times 10^{-14}$ i.e., a 25\% decrease in the disc luminosity leads to a 500\% decrease in the mass flux. For $\Gamma \gtrsim \Gamma_{\rm{min}}$ the mass flux is a strong function of luminosity and our factor of $\sim 5$ disagreement of the mass flux with the 2D simulations is therefore less surprising. 

The other noticable discrepancy is the fast-stream velocity of Model D. As the stellar luminosity is increased, the wind opening angle decreases and the fast-stream narrows. Along the narrow angular slice over which density varies from polar funnel density $\rho \sim 10^{-22} \rm{g \ cm^{-3}}$, the density floor, to wind density $\rho \sim 10^{-13} \rm{g \ cm^{-3}}$ the wind velocity decreases from 4500 to 3500 km/s. The fast-stream velocity is thus very sensitive to where the velocity is measured. We note that the total mass flux agrees to $\sim 4 \%$ so the decrepency in outflow velocity is due to how we are defining it.

\subsection{Local Deviations}

\begin{figure}
    \centering
        \includegraphics[width=0.5\textwidth]{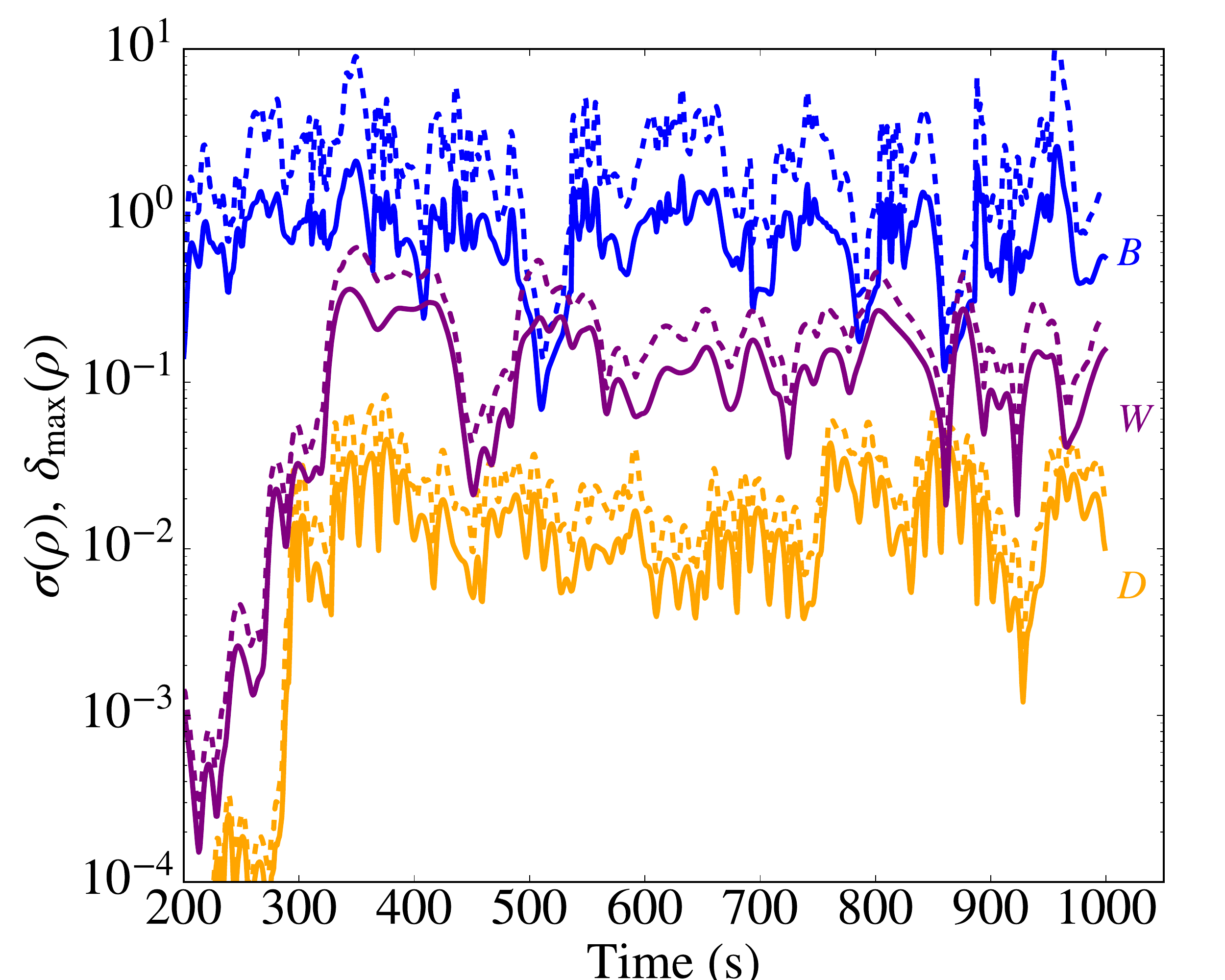} 
    \caption{Relative standard deviation $\sigma$ (solid line) and relative maximum deviation $\delta_{\rm{max}}$ (dashed) at the disc, base and wind points as a function of time for Model A.}
\label{fig:deviation}
\end{figure}

To quantify departures from axisymmetry, we define the relative standard deviation
\begin{equation}
\sigma\left( \rho \right) = \frac{1}{\bar{\rho}} \sqrt{\sum_{k = 0}^{N_{\phi}}\left( \rho_k - \bar{\rho}\right)^2}.
\label{eq:stddev}
\end{equation}
This metric quantifies how density at fixed $(r,\theta)$ deviates, on average, in the $\phi$ direction. We also define the relative maximum deviation
\begin{equation}
\delta_{\rm{max}}\left( \rho \right) = \frac{1}{\bar{\rho}} \rm{max}\Big| \rho_k - \bar{\rho} \Big|.
\end{equation}
This provides a measure of the largest deviation from axisymmetry at a fixed $(r,\theta)$. Because emission scales like $\rho^2$, we consider over(under)-dense regions observationally relevant. To potentially emit $\sim 10$ times more (less) than the azimuthal average, the density must differ by a factor of about three from the azimuthal average. This motivated us to refer to regions with $\rho_k \geq 3 \bar{\rho}$ as over-dense and those with $\rho_k \leq \bar{\rho}/3$ as under-dense. 

In Fig. \ref{fig:deviation}, we plot the relative standard deviation and relative maximum deviation at the disc, base and wind points as a function of time for Model A. Non-axisymmetries are largest at the base, where $\sigma \approx 1$, compared to $\sigma \approx 10^{-1}$ in the wind and $\sigma \approx 10^{-2}$ in the disc. We see that oscillations in $\sigma$ and $\delta_{\rm{max}}$ follow each other, suggesting that variations in the mean standard deviation are primarily due to variations in the maximum deviation i.e. the most prominent clump. We notice several important differences from the perturbed disc model DP18p (Fig 4 from DP18). Deviations are larger at each point, with roughly a factor of 2 difference in the disc and the base and a factor of 10 in the wind. These perturbations take $\sim 300$ s to form in the disc, unlike when the disc is perturbed and the deviations begin at t = 0. At the base, the relative and maximum deviations still correlate with each other. The perturbed flows had $\delta_{\rm{max}} \sim 2 \sigma$, whereas the Fig \ref{fig:deviation} shows $\delta_{\rm{max}} \sim 5 \sigma$, indicating that non-axisymmetric features are more pronounced.    

\subsection{Autocorrelation Length}

\begin{figure*}
    \centering 
        \includegraphics[width=0.45\textwidth]{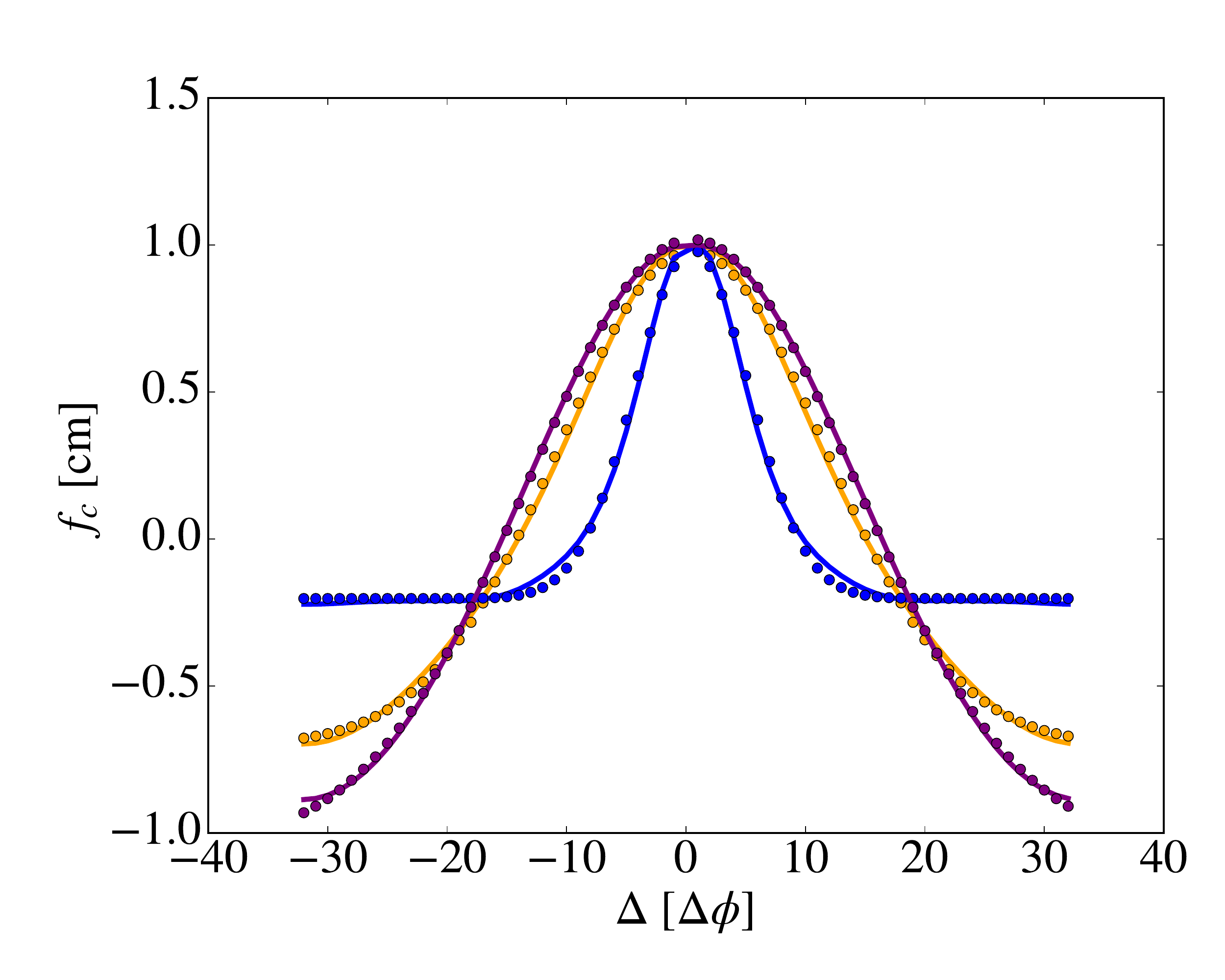} 
        \includegraphics[width=0.45\textwidth]{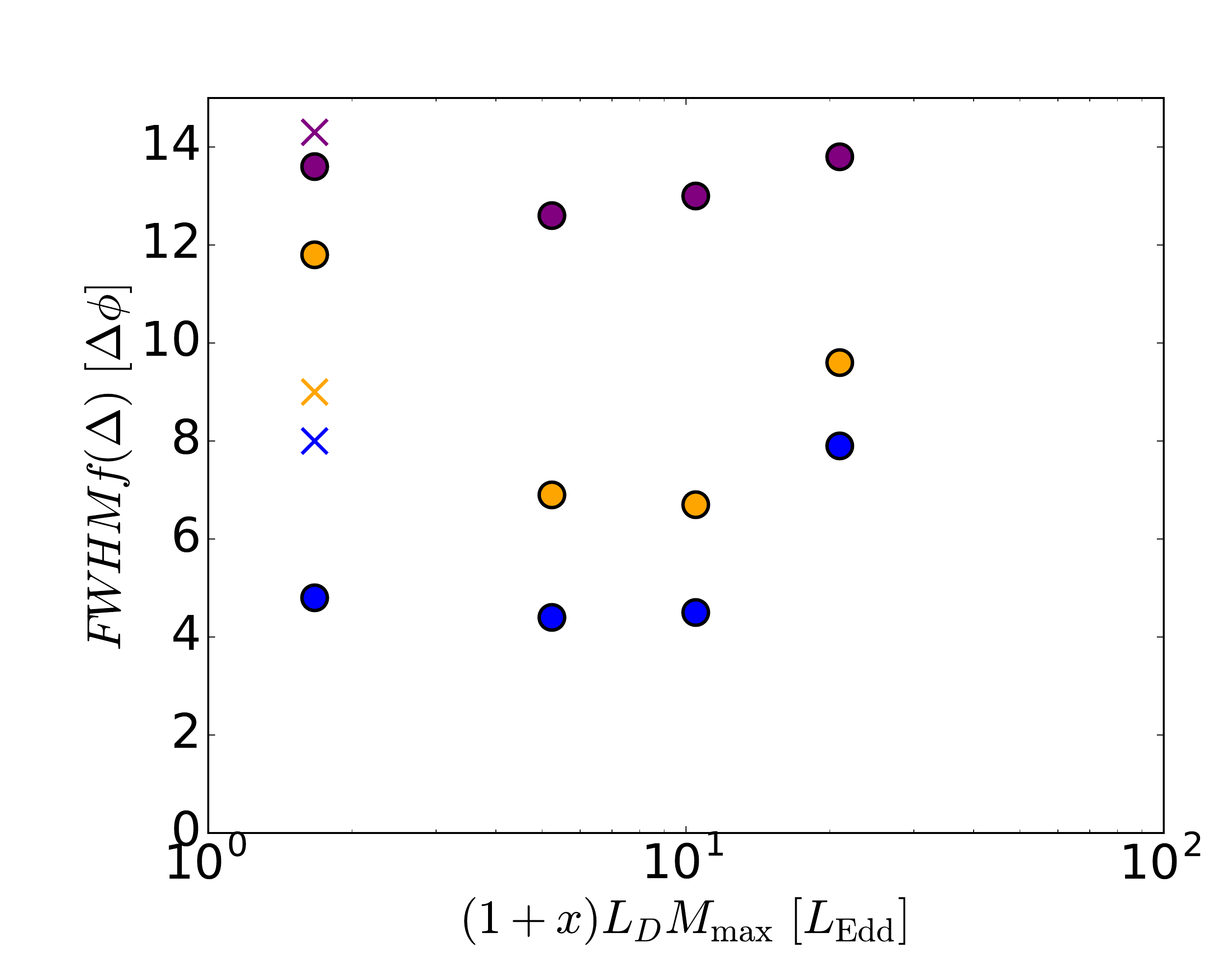} 
    \caption{\textit{Left -} Density autocorrelation for disc (orange points), base (blue points) and wind (purple points) for Model A and Gaussian fit (solid line) to determine angular correlation length.    \textit{Right -} Angular correlation length for each model (solid points) and the perturbed disc model from DP18p (crosses). Clumps are well resolved at the base of the wind whereas in the wind they are of order the wedge size.}
\label{fig:correlation1}
\end{figure*}
\begin{figure*}
    \centering
        \includegraphics[width=\textwidth]{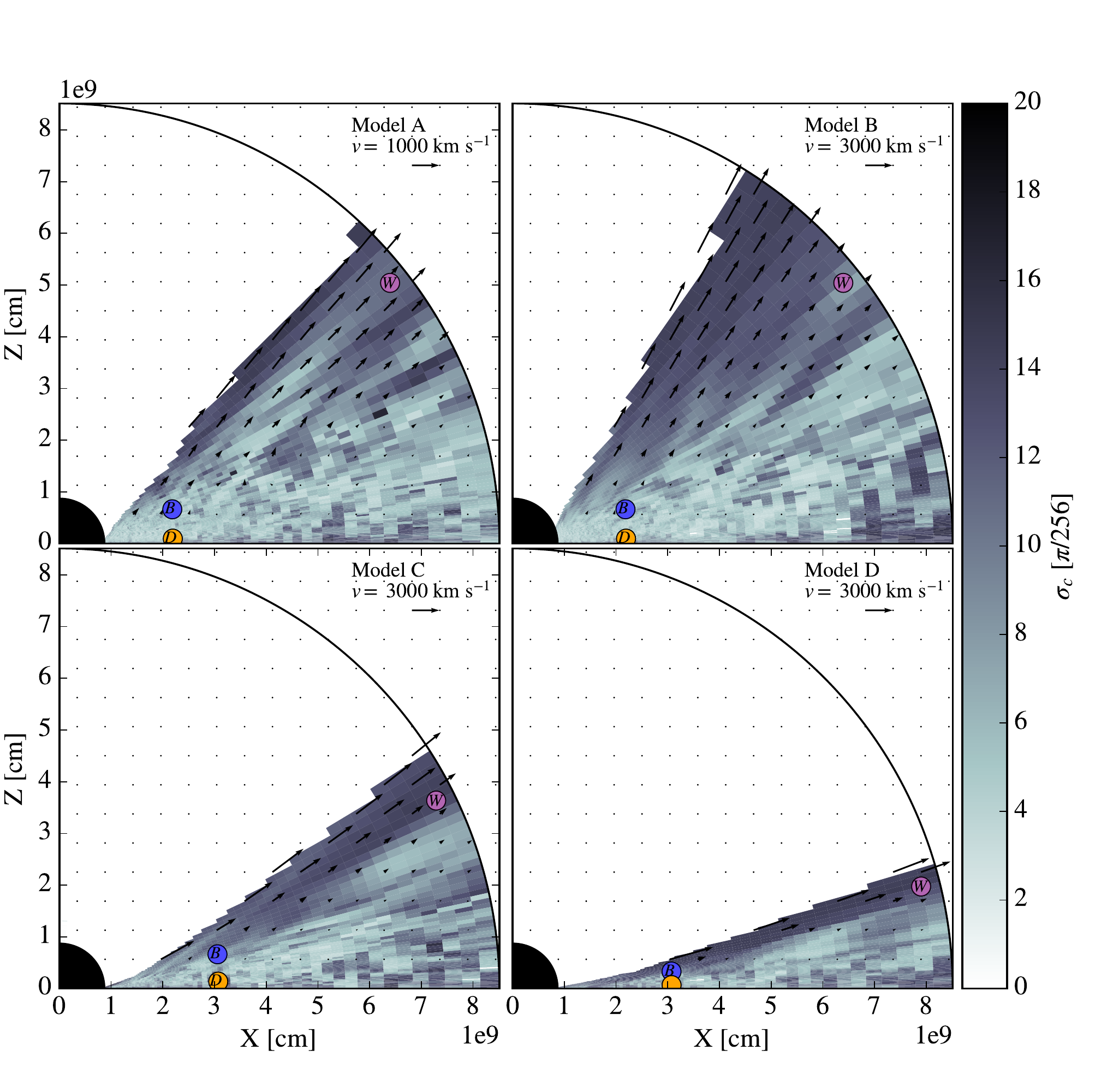}  
    \caption{Angular correlation length $\sigma_c$ for each model averaged over $900 \ \rm{s} \leq t \leq 1000 \ \rm{s}$ from fitting density autocorrelation function. Density structures on scales smaller than the wedge size are restricted to the base of the wind, where the flow is slower whereas the fast-stream has no axial structure.}
\label{fig:correlation2}
\end{figure*}

To estimate the typical azimuthal size of clumps we consider the density autocorrelation in the $\phi$ direction
\begin{equation}
f_c(\Delta) = \frac{\sum_t \sum_k \left( \rho_{k,t} - \langle \rho \rangle \right) \left( \rho_{k-\Delta,t} - \langle \rho \rangle \right)}{\sum_t \sum_k \left( \rho_{k,t} - \langle \rho \rangle \right)^2},
\end{equation} 
where k indexes cells in the $\phi$ direction, t indexes time, $\langle \rho \rangle$ is the time-average over k and the size is measured in units of the $\phi$ grid spacing $\Delta$. We calculate the density autocorrelation over the time range $900 \ \rm{s} \leq t \leq 1000 \ \rm{s} $ for points in the wind where $\rho > 10^{-17} \rm{g \ cm^{-3}}$. In Fig \ref{fig:correlation1}, we show the density autocorrelation for the disc, base and wind points for Model A (colored points) and fit them to Gaussian distributions (colored lines). We use the standard deviation of these fits $\sigma_c$ to define the angular correlation size of the clumps as the FWHM or $\Delta \phi_c \approx 2.35 \sigma_c$. The angular correlation length is related to a physical length scale via $l_c = r \Delta \phi_c$. 

In a periodic system of size L, the correlation length of the $\rm{n^{th}}$ mode is a perfect sinusoid i.e. $f_c(\Delta) = \cos(2n\pi\Delta/L)$. We see that this is nearly the case for modes in the wind where $f(\pm \pi/8) \approx -1$ and to a lesser extent also for modes in the disc. In the base of the wind, we resolve density features with width much smaller than the domain size, $\Delta \phi_c \ll \pi/4$. This corresponds to a physical length scale of $l_c \approx 3 \times 10^{8} \ \rm{cm}$. This is much larger than the Sobolev length in the radial direction $l_{\rm{sob}} \sim v_{\rm{th}} \ dr/dv_r \approx 1 \times 10^{7} \ \rm{cm}$.  

Extending this procedure to the entire wind, in Fig \ref{fig:correlation2}, we plot the angular correlation length $\sigma_c$ for each wind model (upper panels). The fast-stream is smooth with an angular correlation length comparable to the wedge size.  In the slower part of the flow some regions have $\sigma_c \lesssim 10$ and others $\sigma_c > 10$, as expected from the distribution of over-dense and under-dense clumps in the $r-\theta$ plane. By comparison, the base of the wind has $\sigma_c \lesssim 4$ rather uniformally. For $\sigma_c \lesssim 4$, density features are well within the resolution of our simulation with $N_{\phi} = 64$. We conclude that clumps at the base of the wind are a common feature in line driven winds.

We notice two distinctions between our models with different $x$. The first is a consequence of primarily stellar driven models (large $x$) to be more time-independent. This leads to less clumpy outflows. The second is that increasing the stellar contribution decreases the opening angle of the wind and hence reduces the physical extent of the wind base where clumps are largely restricted to.     

In the right panel of Fig \ref{fig:correlation1}, we plot the angular correlation length for the disc, base and wind points in each model (points) and for the perturbed model in DP17 (crosses). We do not see any relationship between the size of clumps and the relative disc and stellar luminosities. This suggests that changing relative stellar and disc luminosity can change the size of the base and wind regions but does not strongly affect the azimuthal clump size.

\subsection{Velocity Dispersion}
\begin{figure*}
    \centering
        \includegraphics[width=\textwidth]{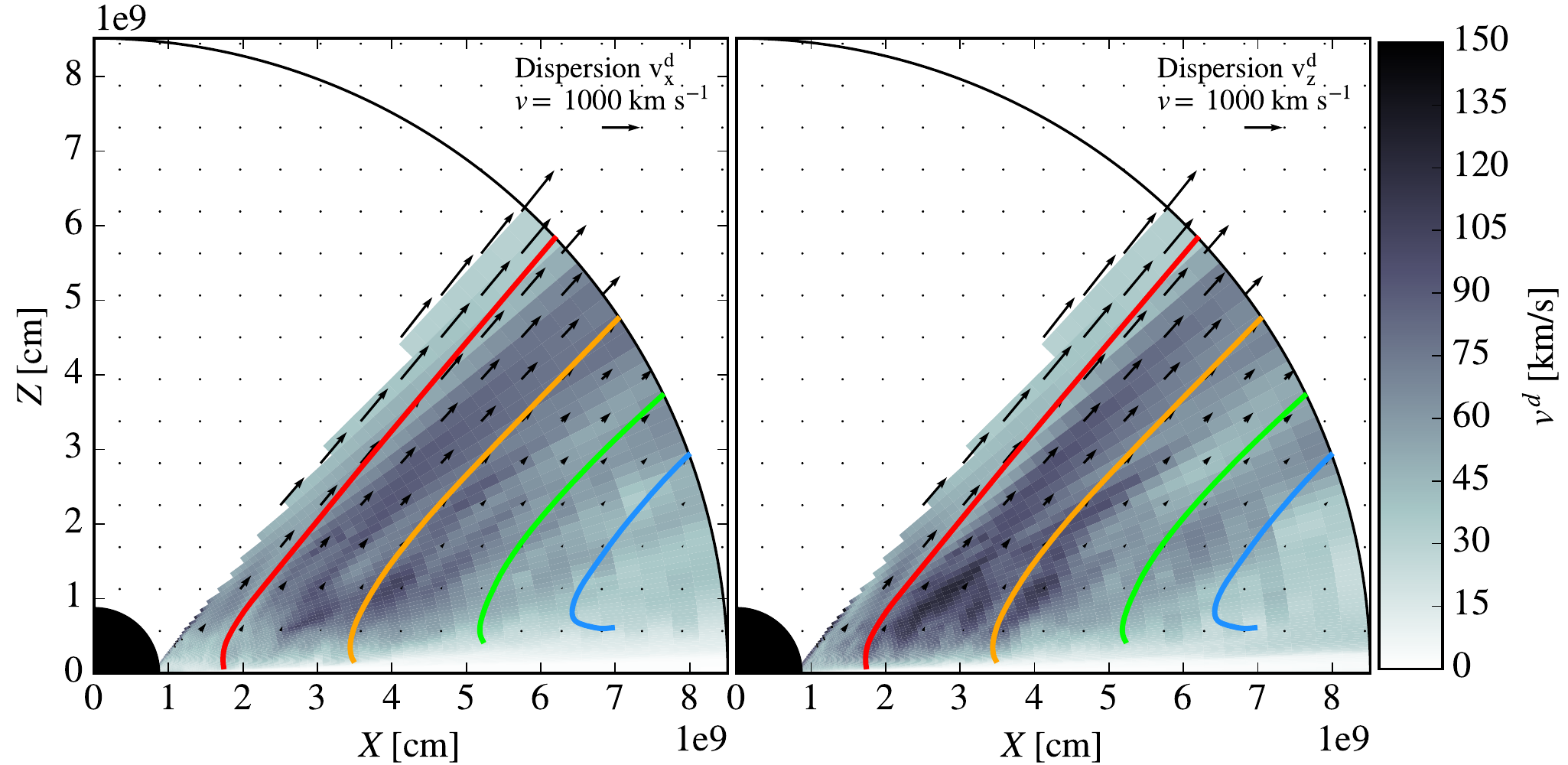}
        \includegraphics[width=\textwidth]{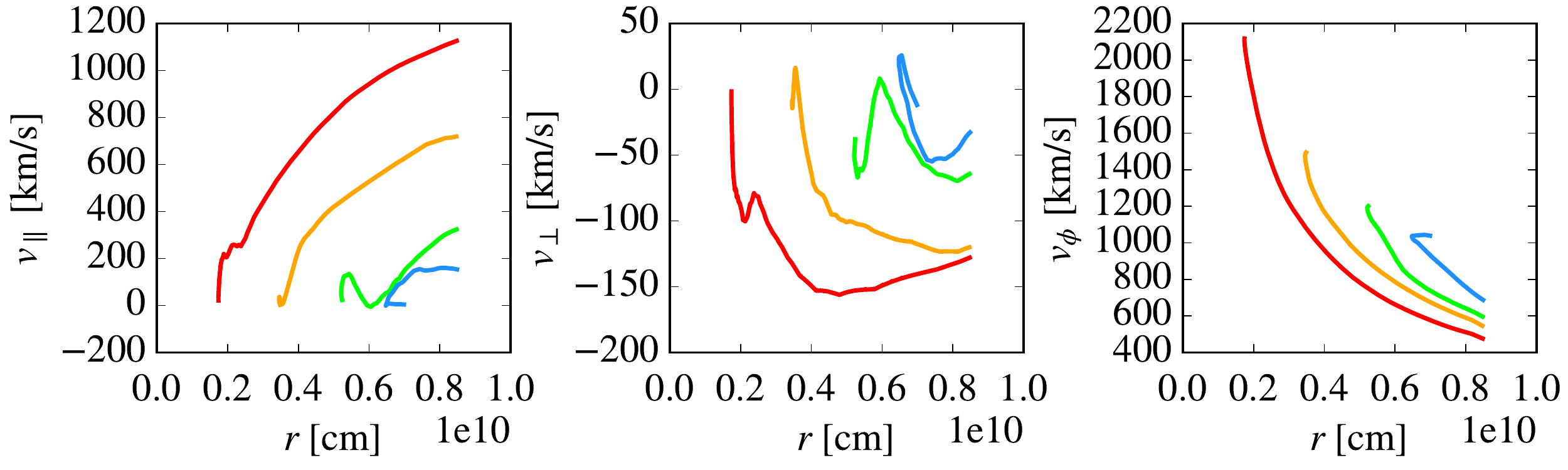} 
        \includegraphics[width=\textwidth]{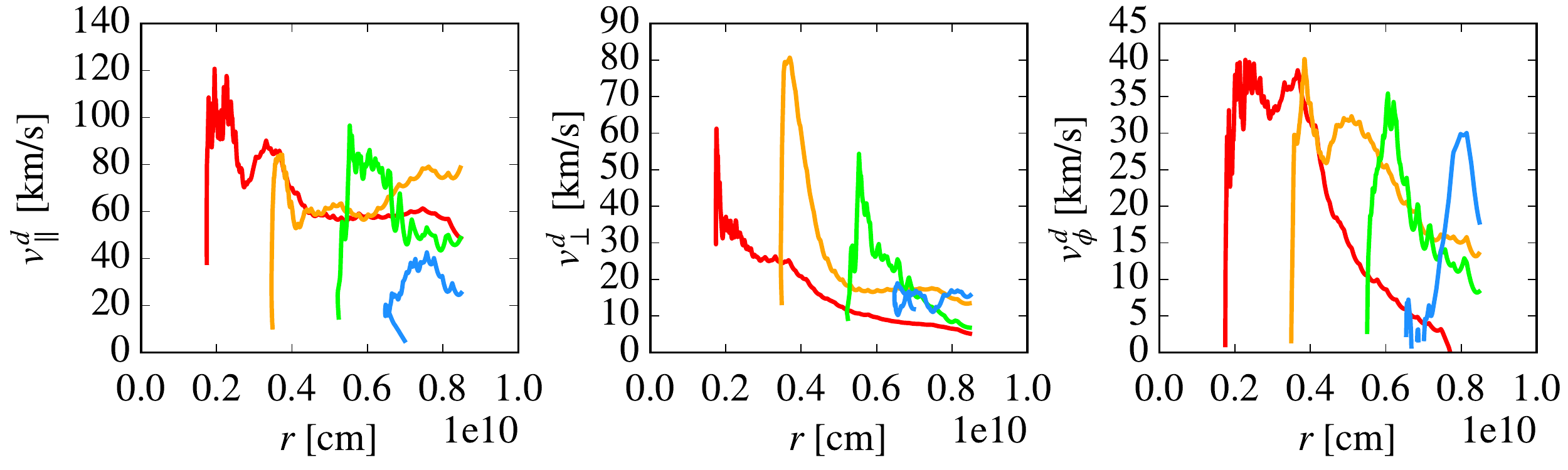}
    \caption{\textit{Top -} Velocity dispersion contours $v^d_x$ and $v^d_z$  in the $\hat{x}$ and $\hat{z}$ directions respectively for Model A averaged over $900 \ \rm{s} \leq t \leq 1000 \ \rm{s}$. We indicate the location of four representative streamlines (red, orange, green and blue lines) \textit{Middle -} Axially and time-averaged velocity parallel $v_{\parallel}$, perpendicular $v_{\perp}$ and in the azimuthal direction $v_{\phi}$ along the correspondingly colored streamlines. \textit{Bottom -} Velocity dispersion parallel $v^d_{\parallel}$, perpendicular $v^d_{\perp}$ and in the azimuthal direction $v^d_{\phi}$ along the colored streamlines. }
\label{fig:velocity_stream}
\end{figure*}

For velocity component in the $\hat{n}$ direction, we define the velocity dispersion
\begin{equation}
v^{\rm{d}}_{\hat{n}} = \sqrt{\langle v_{\hat{n}}^2 \rangle - \langle v_{\hat{n}} \rangle^2}. 
\end{equation}
For fixed coordinate directions, say $\hat{x}$ and $\hat{z}$ this time average is straightforward. However, to facilitate comparison with SOP17 we break up the velocity into components \emph{parallel} and \emph{perpendicular} to streamlines, $v_{\parallel}$ and  $v_{\perp}$ respectively. In this case, we compute streamlines for a time-averaged wind and compute the velocity dispersion relative to this fixed streamline.  
 
In Fig \ref{fig:velocity_stream}, we plot the velocity dispersion $v^d_{x}$ (upper left panel) and $v^d_z$ (upper right panel) in the $\hat{x}$ and $\hat{z}$ directions respectively for Model A averaged over $900 \ \rm{s} \leq t \leq 1000 \ \rm{s}$. We overplot four representative streamlines for the time averaged wind (colored lines) and plot the velocity (middle panel) and velocity dispersion (lower panel) for each streamline, in their respective color, in the parallel, perpendicular and axial directions.

Near the outer boundary, the velocity dispersion is greatest in the fast-stream where $v^d_x \approx v^d_z \sim 100 \ \rm{km/s}$ which corresponds to roughly $\mathbf{v}^d \approx 5-10\% \ \mathbf{v}$. In the slower parts of the flow, the dispersion is more modest with $v^d_x \approx v^d_z \sim 30 \ \rm{km/s}$ and is nearly vanishing in the disc. When the flow is more vertical (Model B) dispersion is greater in $v^d_{z}$ whereas when the flow is more radial (Models C and D) the dispersion in $v^d_{x}$ is greatest. 

Along streamlines $v^d_{\parallel} \approx 3-10\% \ v_{\parallel}$, where relative dispersion decreases with increasing stellar luminosity. This is expected since these are the least time dependent solutions. We find similar relative dispersion for $v^d_{\perp}$, with the exception of Model D where $v^d_{\perp} \approx 50 \% \ v_{\perp}$. This is the most stationary model and therefore the absolute velocity $v_{\perp}$ is small.  

For steamlines in the fast-stream (red and orange) velocity dispersion near the outer boundary $v^d_{\parallel} \approx 5\% \ v_{\parallel}$. However at the base $v^d_{\parallel} \approx 50\% \ v_{\parallel}$. If we think of the dispersion velocity as a sort of turbulent motion, this suggests that line broadening may be due to motion of clumps at the base of the wind, where the flow is less well ordered.       

The dispersion relation is different from that generated by the LDI. SOP17 find (their Fig 6) that $v^d_{\parallel}$ is small at the base of the wind but steadilly increases. We find $v^d_{\parallel}$ oscillates, but there is no monotonically increasing trend and at larger radii remains mostly constant. We find $v^d_{\perp}$ peaks at small radii before settling to a constant value similar to SOP17 that find it quickly settles to a constant value. These findings suggest that parallel to the streamline, the LDI is responsible for a larger velocity dispersion than geometric effects due to the wind clumping. This is particularly true at larger radii in the fastest part of the flow where our winds become less clumpy.

\subsection{Clumpiness \& Column Density}

\begin{figure}
    \centering
        \includegraphics[width=0.5\textwidth]{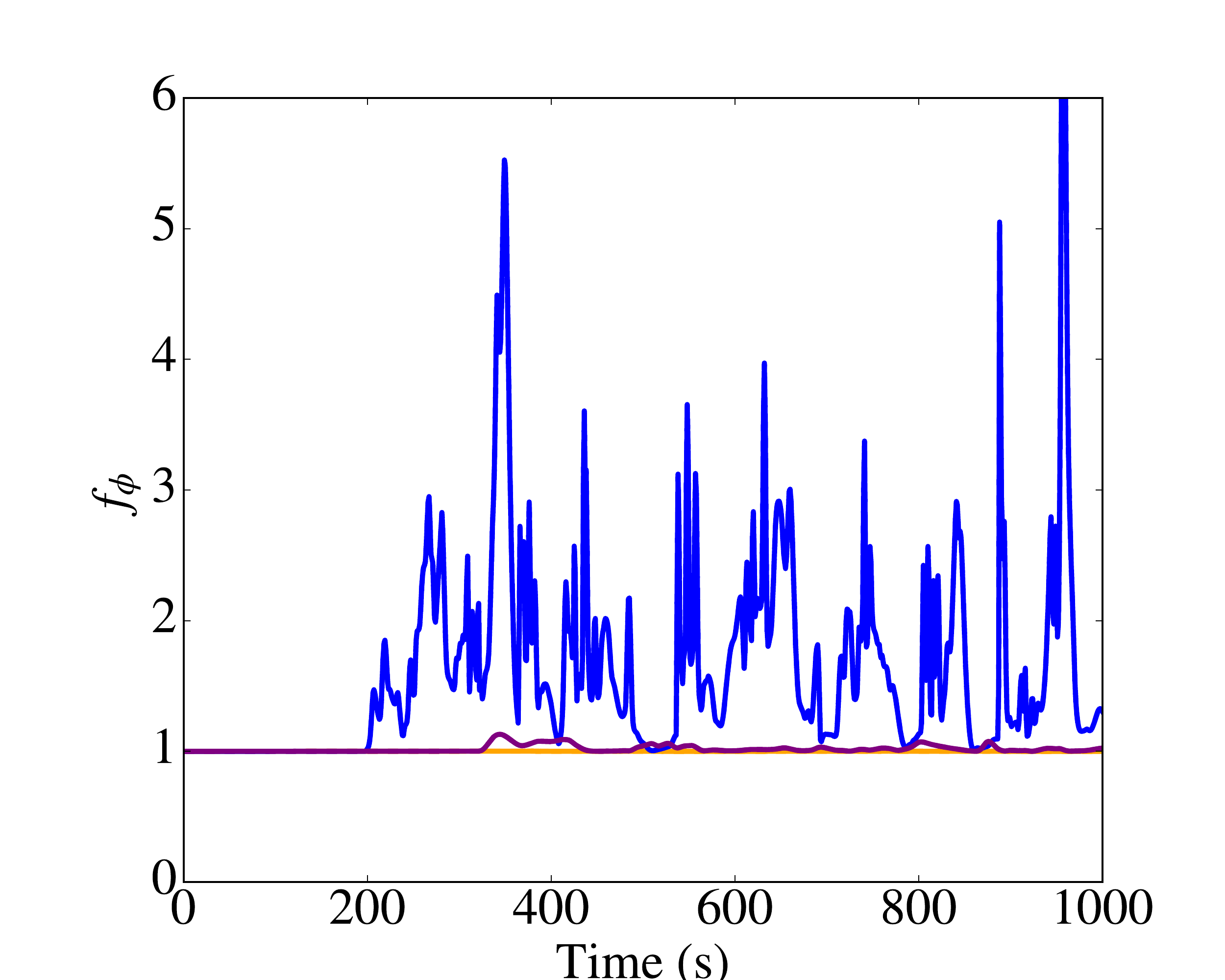} 
    \caption{Azimuthal clumping $f_{\phi}$ for Model A as a function of time. Clumpiness spikes are dominated in the base and are approximately 50\% larger and 5 times broader than the same distribution for the perturbed disc in DP18.}
\label{fig:clumpiness_phi}
\end{figure}

\begin{figure*}
    \centering
        \includegraphics[width=\textwidth]{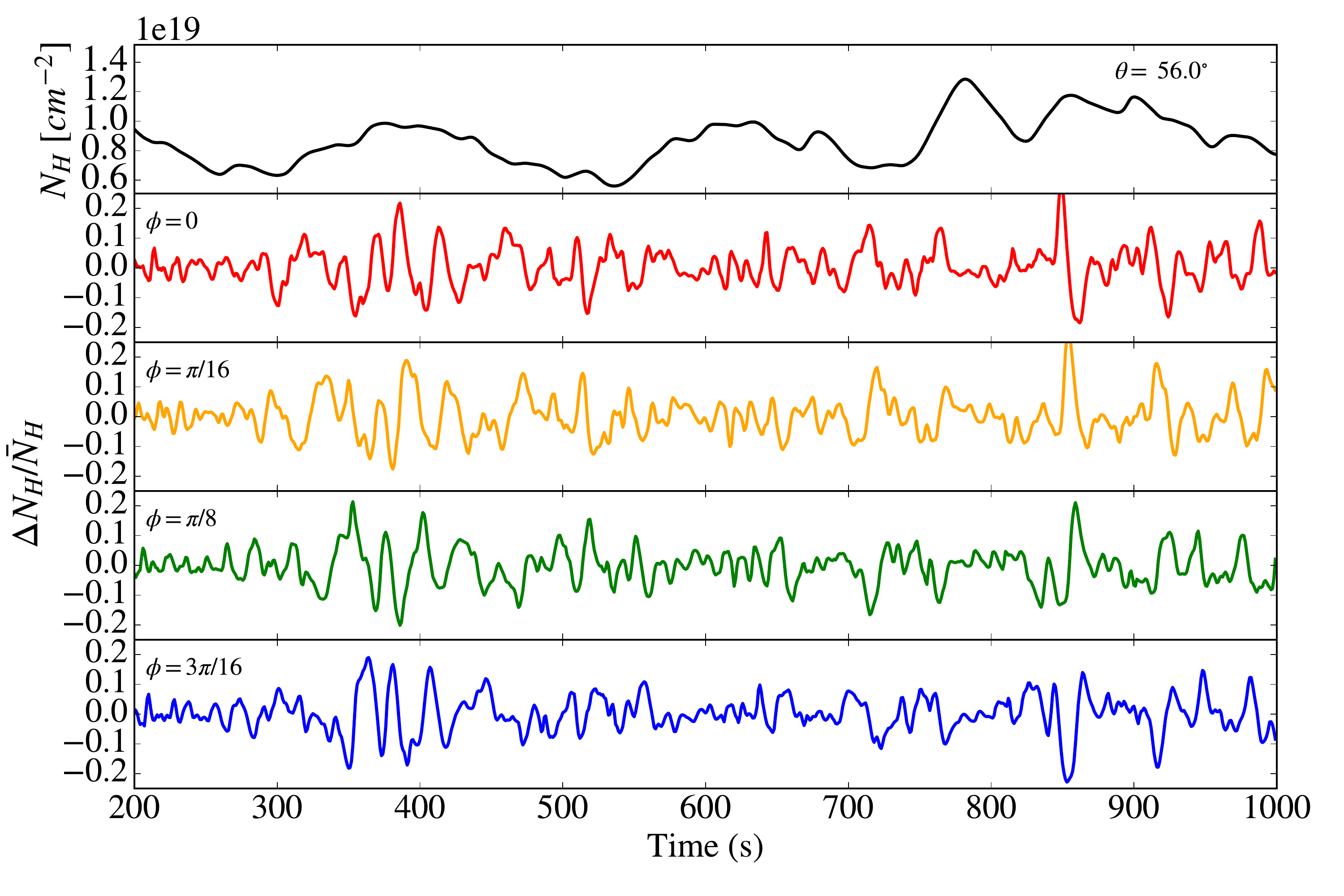} 
    \caption{\textit{Top panel - }$\phi-$averaged column density $N_H$ at $\theta = 56^{\circ}$ as a function of time. \textit{Lower panels - } Column density along $\phi = $ 0 (red line), $\pi/16$ (orange line), $\pi/8$ (green line), $3\pi/16$ (blue line). Variations of $\sim 20 \%$ in the column density along fixed $\phi$-sightlines are primarily due to density features at the base of the wind.}
\label{fig:column_density}
\end{figure*}

To quantify and identify the origin of the non-axisymmetry we measure the clumping factor defined as 
\begin{equation}
f_{\phi}(r,\theta) = \frac{ \overline{\rho^2}}{ \overline{\rho} ^2}.
\end{equation}
This is a useful metric because $f_{\phi} \equiv 1$ for axisymmetric simulations and is therefore a local measurement of deviations from axisymmetry. 

In Fig. \ref{fig:clumpiness_phi}, we plot the azimuthal clumping $f_{\phi}$ for Model A as a function of time at the disc, base and wind points. Clumping is largely restricted to the base of the wind, where $1 < f_{\phi} < 6$ and peaks have width $\sim 100$ s. The perturbed disc case DP18p also had departures from axisymmetry ($f_{\phi} \neq 1$) primarily restricted to the base of the wind, but had less clumping with $1 < f_{\phi} < 4$ (Fig 5 in DP18) . Also the spikes were shorter in duration lasting $\sim 20$ s. Models B and C have a similar distribution of $f_{\phi}$ while for Model D $f_{\phi} \approx 1$, consistent with our other metrics that characterize the clumping in Model D as the least significant.

The formation of azimuthal density structures affects the column density 
\begin{equation}
N_H(\theta,\phi) = \int_{r_{*}}^{10r_{*}} \rho(\theta,\phi) \,dr.
\end{equation}
along radial lines of sight. Though these are not directly observable (we cannot measure column density along a fixed $\phi$ sightline), they may affect the optical depth. We also define the $\phi-$averaged column density
\begin{equation}
\bar{N}_H(\theta) = \frac{4}{\pi}\int_0^{\pi/4} N_H(\theta,\phi') \, d\phi'.
\end{equation}
In Fig \ref{fig:column_density}, we plot the $\phi-$averaged column density along the sightline at $\theta = 56^{\circ}$ as a function of time (top panel) and the normalized column density deviation $(N_H - \bar{N}_H)/\bar{N}_H$ at fixed $\phi$ (lower panels) for $\phi = 0$ (red), $\pi/16$ (orange), $\pi/8$ (green), $3\pi/16$ (blue) for Model A. The deviations in column density as a function of $\phi$ can be thought of as a radially integrated measure of clumpiness along a line of sight. We see that $\bar{N}_H \sim 10^{19} \rm{cm}^{-2} \ll 10^{24} \rm{cm}^{-2}$ so our assumption that the wind is optically thin in the continuum is justified.

Along fixed $\phi$ sightlines, the column density $N_H(\phi)$ varies $\sim 20\%$, comparable to variations in the $\phi-$averaged deviations $\bar{N}_H$ in time. Deviations in the column density along fixed $\phi$ can be interpreted as density features or clumps orbiting in the wind and crossing the line of sight. As an example, we see a peak in the column density $N_H(0)$ at t = 844 s, and a similar sized peak in the column density $N_H(3\pi/16)$ at t = 859 s. This corresponds to an angular velocity $\Omega \approx 0.04 \ \rm{s}^{-1}$, the Keplerian angular velocity at r = $3.7 \times 10^{9}$ cm. This radius is at the base of the wind where the wind is most clumpy. Interestingly, this is also the approximate position of clumps for the perturbed wind in DP18, which were located by tracking variations in the clumpiness (see their Fig 6).        

\subsection{Fourier Transform}

\begin{table}
\begin{center}
    \begin{tabular}{| l | c |l | l | l |}
    \hline \hline
Model 		& Variable 	& \multicolumn{3}{c}{Modes} \\ 
		& 		&Disc	& Base & Wind \\ \hline \hline
\multirow{4}{*}{DP18u}& $\hat{\rho}$			& 1,3		& 1,..,12	& 1		\\ 
			& $\hat{v}_{r}$			& 1		& 1		& 1		\\
			& $\hat{v}_{\theta}$		& 1,2,3,4	& 1		& 1		\\ 
			& $\hat{v}_{\phi}$		& 1		& 1,2,3		& 1		\\ \hline
\multirow{4}{*}{DP18p}& $\hat{\rho}$			& 1,3,7,13,17	& 1,..,12	& 1		\\ 
			& $\hat{v}_{r}$			& 1,9,11	& 1,6,7,8,12	& 1		\\
			& $\hat{v}_{\theta}$		& 1,2,8,12,17	& 1,7,12	& 1		\\ 
			& $\hat{v}_{\phi}$		& 1,9,11	& 1,6,7,8,12	& 1		\\ \hline \hline
\multirow{4}{*}{A}	& $\hat{\rho}$			& 1,3,7,13,17	& 1,..,12	& 1		\\ 
			& $\hat{v}_{r}$			& 1,9,11	& 1,6,7,8,12	& 1		\\
			& $\hat{v}_{\theta}$		& 1,2,8,12	& 1,6,7		& 1		\\ 
			& $\hat{v}_{\phi}$		& 1,9,11	& 1,6,7,8	& 1		\\ \hline
\multirow{4}{*}{B}	& $\hat{\rho}$			& 1,3,7,13,17	& 1,3,..,12	& 1		\\ 
			& $\hat{v}_{r}$			& 1,9,11	& 1,5,6,7,8	& 1		\\
			& $\hat{v}_{\theta}$		& 1,2,8,12	& 1,6,7,8	& 1		\\ 
			& $\hat{v}_{\phi}$		& 1,9,11	& 1,6,7,8	& 1		\\ \hline
\multirow{4}{*}{C}	& $\hat{\rho}$			& 1		& 1,4,7,8	& 1		\\ 
			& $\hat{v}_{r}$			& 1,6,11,13	& 1,4,7,8,9,11	& 1		\\
			& $\hat{v}_{\theta}$		& 1,5,8,11	& 1,4,7,8,9,11	& 1		\\ 
			& $\hat{v}_{\phi}$		& 1,6		& 1,4,7,9,11	& 1		\\ \hline
\multirow{4}{*}{D} 	& $\hat{\rho}$			& 1,5,8,11	& 1,5,6,9,11	& 1		\\ 
			& $\hat{v}_{r}$			& 1,6,7,11,12,13& 1		& 1		\\
			& $\hat{v}_{\theta}$		& 1,5,7,11	& 1,4		& 1		\\ 
			& $\hat{v}_{\phi}$		& 1,5,6,7,12,13	& 1,4		& 1		\\ \hline \hline
    \end{tabular}
\end{center}
\caption{Summary of Fourier modes for each model. Models with the same line driving (DP18p and A) but different mechanism for producing non-axisymmetries exhibit the same modes, as do models with only disc disc radiation (Models A and B). Axisymmetric (DP18u) and nearly axisymmetric models (D) exhibit the least number of modes.}
\label{table:modes}
\end{table}

\begin{figure*}
    \centering
        \includegraphics[width=\textwidth]{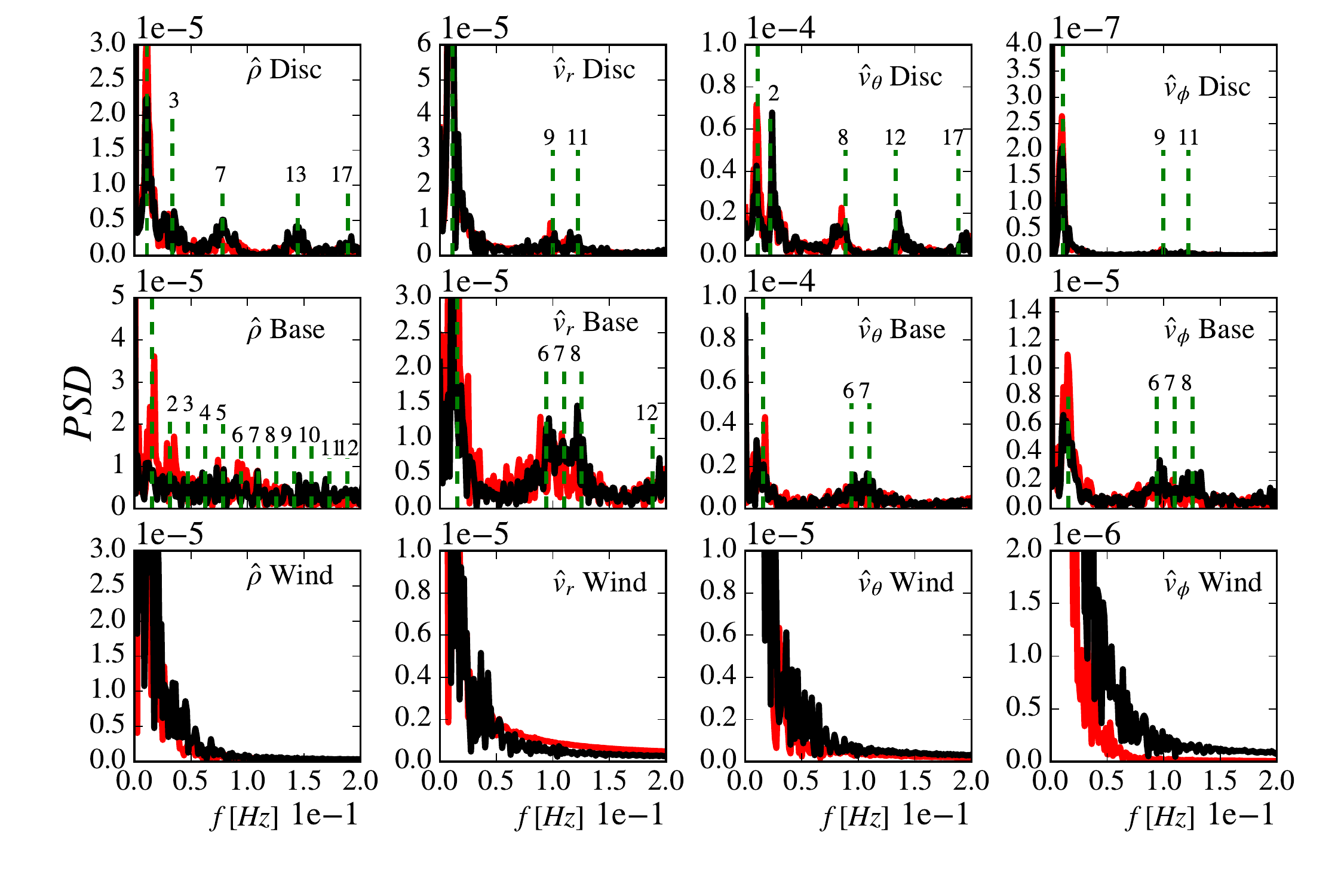}
    \caption{Fourier modes for Model A (black lines) and DP18p (red lines) for the disc, base and wind points. Both models exhibit the same modes, suggesting that the mechanism breaking axisymmetry (an initial disc pertubation in the case of DP18p and geometric line driving effects in the case of Model A) has no effect on the modes of the solution. We note that the approximate power in Fourier modes is different for each of the dynamical variables.}
\label{fig:fourier}
\end{figure*}

To further investigate the time variability of the non-axisymmetry, we consider the Fourier transform, 
\begin{equation}
\hat{\rho}(\rho,\theta,\phi,\omega) = \int_{-\infty}^{\infty} \rho(\rho,\theta,\phi,t) \, e^{-2 \pi i \omega t} \ dt,
\end{equation}
where we apply a standard fast fourier transform (FFT) algorithm on our data. To eliminate any effects due to transient modes associated with the initial wind launching, we consider times $200 \ \rm{s} \leq t \leq 1000$ s, as the time during which an outflow is established.

In Fig. \ref{fig:fourier}, we plot the power spectrum distribution (PSD) of the density $\hat{\rho}$ and the velocity components $\hat{v}_r$, $\hat{v}_{\theta}$ and $\hat{v}_{\phi}$ for Model A (black lines) and the perturbed disc model DP18p (red lines). We find both models excite the same Fourier modes. We see some differences in the spectra at lower wavenumber where at the base of the wind, DP18p has slightly higher power in the m = 1 and 2 modes. Line driving is not expected to be dominant in the disc, therefore we expect any modes to be due to disc oscillations, independent on the wind driving mechanism.  This is consistent with the results in DP18 where we found that different disc pertubations (different wavenumber, different amplitude) excited the same modes in the disc and base of the wind.  

In Table \ref{table:modes}, we summarize which modes are excited in each of the models. In Models A and B, because of the similar wind geometry, we computed Fourier modes at the same locations in the flow. We see that despite a factor of 3 increase in the disc luminosity, only a few new modes are excited. For Models C and D, the comparison is more difficult because changes in the wind geometry force us to calculate modes at different locations in the flow and necessarily alters which modes are excited.    

In DP18, we showed that breaking axisymmetry excited additional modes evidenced by the larger number of excited modes in DP18p than DP18u. Similarly we find that the most axisymmetric wind (Model D) exhibits the least number of Fourier modes.  

\section{Discussion}
\label{sec:discussion}

In our optically thin treatment, the inner and outer parts of the flow are only coupled through the flow of gas. Steep velocity gradients tend to smooth out non-axisymmetries, making it so that the faster, outer flow is smooth. If we relax the optically thin assumption, then inner and outer flow are coupled through the radiation field. This will affect the wind by decreasing the available driving flux and altering the ionization state of the gas. Accounting for this first effect requires treating the radiation transfer of continuum photons through the wind. The clumps seen in our simulations are due to having extended radiation sources, both disc and star. We expect that extinction due to clumps will be modest for extended sources since the intensity of radiation will be averaged over all directions. The effects of clumps may be much more significant in altering the ionization state of the gas. Ionizing radiation may be assumed to be emitted from a central point source (in AGN for example). In this case, ionization state of the gas depends on the gas density along a single line of sight and can be affected by single over/under densities. We find variations of $\sim 20 \%$ in the column density along sightlines with different $\phi$ and can expect similar variations in the ionization state. This may be particularly siginificant for low luminosity systems where a sharp turnover in the mass flux has been observed.

Density clumps are restricted to the base of the wind where the flow is slow. However, the velocity dispersion is found to be $\mathbf{v}^{d} \sim 10\% \ \mathbf{v}$. For winds with velocity $|\mathbf{v}| \sim 1000 \ \rm{km/s}$ this is a dispersion of $\sim 100 \ \rm{km/s} \gg c_s = 14 \ \rm{km/s}$. We expect line broadening from non-axisymmetries to be more important than due to thermal broadening.

We find that geometric effects generate clumps at the base of the wind on length scales $l_{c} \approx 10^{8} \ \rm{cm} \lesssim R_*$. SOP17 found the the LDI generates clumps on order the Sobolev length $l_{\rm{sob}} \approx 10^{7} \ \rm{cm}$. This suggests that fully treating the radiative transfer will produce clumps over a hierarchy of scales $l_{\rm{sob}} \lesssim l_c \lesssim R_*$. That is, we expect the LDI mechanism for increasing clumps to further increase the clumpiness of the flow. However, it is unclear how Sobolev scale modes couple to super-Sobolev scale modes.

We used a simplified treatment of the hydrodynamics by assuming the flow is isothermal. Allowing the gas to cool is important as non-axisymmetries may grow due to thermal instability. Fully resolving the thermal instability requires much higher resolution than currently available. Simulations of clouds in the context of AGN required $\sim 10$ grids per density scale height in the cloud/medium interface. In our case we see $\sim 10$ clouds along a sightline, suggesting the need for $\sim 100$ times the resolution to fully resolve cloud dynamics. However, allowing gas to be heated by the central source is still expected to allow non-axisymmetries to grow.

We have confirmed in 3D the findings of PSD99 in 2D axisymmetric simulations that line driven disc winds are time-dependent when disc luminosity is greater than or comparable to stellar luminosity. Further we have shown that non-axisymmetries generated using the full velocity gradient (the full-Q method of this work) are similar to those generated when we approximate the velocity gradient (the dv/dz approximation of DP18) but initally perturb the disc. This latter method is computationally much less expensive, and will be a good first step in exploring the additional physics described above.

\section{Conclusion}
\label{sec:conclusion}

We performed 3D simulations of line driven disc winds and found non-axisymmetric features are generated at the base of the wind and propagate outwards. These features have super-Sobolev length scales, density contrast $\sim$ a few the azimuthal average and a velocity dispersion $\sim$ 10 \% the velocity of the wind. These features may have an observable effect on emission and absorption profiles (which depend on $\rho^2$ and $\rho$ respectively) and on line broadening (which depends on the velocity dispersion).

This work has focused on CV systems, but line driving is a promising mechanism for explaining outflows from AGN where much of the same physics is expected to be relevant. Future work will include additional physical effects to study non-axisymmetric winds. We will include extinction from the ionizing source of radiation, which will more strongly couple the turbulent base of the wind with the smooth, fast-stream via the gas ionization state. We will also include the effects of heating, so that density perturbations can further grow via thermal instability. Carefully resolving the ionization and temperature state will further allow us to calculate the force-multiplier self consistently using a photoionization code rather than through an analytic fitting function as we have done in this work. 

\section*{Acknowledgements}
This work was supported by NASA under ATP grant NNX14AK44G. The authors also acknowledge useful discussions with Tim Waters.

\end{document}